\documentclass[a4paper,11pt]{article}
\usepackage{pos}
\usepackage[prefix, nopostdot, nonumberlist]{glossaries-extra}
\usepackage{xspace}
\usepackage{lineno}
\usepackage{svg}
\usepackage{lipsum}
\usepackage{setspace}
\usepackage{natbib}
\usepackage{enumitem}
\usepackage{wrapfig}
\usepackage{subcaption}
\usepackage{caption}

\renewcommand{\glossarysection}[2][]{}
\newcommand{\mynewabbreviation}[4][]{%
  \newabbreviation[#1]{#2}{#3}{#4}%
  \expandafter\newcommand\csname #2\endcsname{\gls{#2}\xspace}%
  \expandafter\newcommand\csname #2s\endcsname{\glspl{#2}\xspace}%
  \expandafter\newcommand\csname a#2\endcsname{\pgls{#2}\xspace}%
  \expandafter\newcommand\csname an#2\endcsname{\pgls{#2}\xspace}%
}
\setabbreviationstyle[long]{long}
\glsdisablehyper
\GlsXtrEnableEntryCounting{abbreviation}{1}
\input{acronyms.acr}

\newcommand{\degree}{$^\circ$\xspace}
\newcommand{\dmax}{d_\text{max}}
\newcommand{\fluence}{\mathscr{F}}
\newcommand{\vxB}{{\vec v \times \vec B}}
\newcommand{\cel}{c_\text{el}}
\newcommand{\fGS}{f_\text{GS}}

\title{Information Field Theory based Event Reconstruction for Cosmic Ray Radio Detectors }

\author*[a]{Simon Strähnz}
\author[a,b]{Tim Huege}
\author[c]{Torsten Enßlin}
\author[d]{Karen Terveer}
\author[d,e]{Anna Nelles}

\affiliation[a]{Karlsruhe Institute of Technology,\\
  Kaiserstraße 12, 76131 Karlsruhe, Germany}
\affiliation[b]{Astrophysical Institute, Vrije Universiteit Brussel,\\
  Pleinlaan 2, 1050 Brussels, Belgium}
\affiliation[c]{Max-Planck-Institute for Astrophysics,\\
  Karl-Schwarzschild-Str. 1, 85748 Garching b. München, Germany}
\affiliation[d]{Erlangen Centre for Astroparticle Physics,\\
  Friedrich-Alexander-Universit\"at Erlangen-N\"urnberg, 91058 Erlangen, Germany}
\affiliation[e]{Deutsches Elektronen-Synchrotron DESY,\\
  Platanenallee~6, 15738 Zeuthen, Germany}

\emailAdd{simon.straehnz@kit.edu}

\abstract{
Detection of extensive air showers with radio antennas is an appealing technique in cosmic ray physics. However, because of the high level of measurement noise, current reconstruction methods still leave room for improvement. Furthermore, reconstruction efforts typically focus only on a single aspect of the signal, such as the energy fluence or arrival time. Bayesian inference is then a natural choice for a holistic approach to reconstruction, yet, this problem would be ill-posed, since the electric field is a continuous quantity. Information Field Theory provides the solution for this by providing a statistical framework to deal with discretised fields in the continuum limit.

We are currently developing models for this novel approach to reconstructing extensive air showers.
The model described here is based on the best current understanding of the emission mechanisms: It uses parametrisations of the lateral signal strength distribution, charge-excess contribution and spectral shape. Shower-to-shower fluctuations and narrowband RFI are modelled using Gaussian processes. Combined with a detailed detector description, this model can infer not only the electric field, but also the shower geometry, electromagnetic energy and position of shower maximum.

Another big achievement of this approach is its ability to naturally provide uncertainties for the reconstruction, which has been shown to be difficult in more traditional methods. With such an open framework and robust computational methods based in Information Field Theory, it will also be easy to incorporate new insights and additional data, such as timing distributions or particle detector data, in the future. This approach has a high potential to exploit the full information content of a complex detector with rigorous statistical methods, in a way that directly includes domain knowledge.
}

\ConferenceLogo{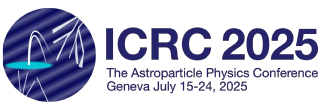}

\FullConference{39th International Cosmic Ray Conference (ICRC2025)\\
 15–24 July 2025\\
Geneva, Switzerland\\}

\begin{document}
\maketitle

\setlength{\parskip}{0cm}
\setlength{\itemsep}{0cm}
\setlength{\bibsep}{0cm}

\section{Introduction}
    Radio detection of air showers has established itself as a cost-effective way of measuring \eass.
    With the deployment of the Auger Radio Detector~\cite{bjarni_this_icrc} it has been shown that a large-scale radio detector can be built and maintained effectively.
    Radio detection is of special interest for \CR research, because it allows a pure and calorimetric measurement of the electromagnetic component of \eass.
    The main disadvantage of this technique is that it is much more susceptible to noise than others.
    Since cosmic ray observatories are generally not built in radio quiet areas, there will always be some narrowband \rfi in every measurement.
    There is also the radio emission from the Milky Way that, while interesting to radio astronomers, must be considered background and noise for \CR research.
    Current methods of dealing with this (see also \cite{simon_this_icrc}) have shown room for improvement.
    A natural way to address this problem is using Bayesian inference, which poses another problem: Unless the emission can be perfectly modelled parametrically, which given the probabilistic nature of air showers seems impossible, the most rigorous way is to use continuous, nonparametric models for the electric field.
    This, however, is an ill-posed problem, as a continuous domain creates an infinitely dimensional latent space.
    The solution to this problem can be found in Information Field Theory~\cite{ensslinIFT}.
    This work establishes a semiparametric way to reconstruct inclined \eass using radio measurements.
    For a similar approach applied to vertical showers, see~\cite{karen_this_icrc}. 
    
    \subsection{Radio Detection}
    Radio detection relies on the coherent radio emission generated during the development of \eass.
    Macroscopically, two processes of emission can be identified, which can be distinguished by their polarisation.
    The ``geomagnetic'' emission is the dominant emission process.
    When the shower develops, charged particles (mainly electrons and positrons) are deflected by the geomagnetic field and thus radiate.
    This emission is polarised in the (negative) $\vxB$ direction (where $\vec{v}$ is the velocity of the primary particle and $\vec{B}$ the geomagnetic field).
    The second, subdominant process is ``charge-excess'' emission.
    During the development of the shower, negative charges accumulate in the shower front leaving positive charges behind in its wake.
    This charge separation also radiates, with a polarisation pointing from the observer to the shower axis.
    For inclined showers the lateral distribution of the geomagnetic emission can be described by a one-dimensional \ldf and the charge excess emission can be parametrised as a fraction of the geomagnetic emission~\cite{schlueter_2023_signal_model}.
    
    \subsection{Information Field Theory}
    \ift studies how information theory can be applied to (physical) fields. Since these fields are continuous, these problems can generally not be solved numerically: When describing a continuous quantity non-parametrically, this creates an infinitely-dimensional latent space, which can neither be represented in memory, nor inferred within a finite time. However, \ift has proven, that it is allowable to treat these inverse problems in the continuum limit, i.e.\ that discretisation does not change the mathematial implications if it is fine enough~\cite{ensslinIFT}. Solving these problems poses another challenge which has been overcome. The continuum limit forces the use of a very high number of degrees of freedom. These cannot be inferred within reasonable time frames with traditional methods such as Markov chain sampling. For this reason, \mgvi has been developed~\cite{knollmüller2020metricgaussianvariationalinference}. It is an algorithm for variational inference, that is finding the posterior distribution by approximating it with a known distribution, using a normal distribution as approximate posterior and the inverse fisher metric as an estimator for its covariance. In cases where the assumption that the posterior is roughly normally distributed does not hold, this method has been extended to \geoVI~\cite{frank2021geometricvariationalinference}, which transforms the posterior distribution into a coordinate system where it is approximately Gaussian and then applies \mgvi.

\section{Forward Model}
At the heart of our method of Bayesian inference stands the forward model, transforming the parameters to be inferred to the data space. For problems as complex as the one at hand, these models become very involved, so first an overview will be given, followed by the details of the implementation.
    \subsection{Overview}
    The forward model is semiparametric, meaning that part of it derives from parametrisation of air shower radio emission, while other parts are nonparametric, i.e.\ continuous. These nonparametric parts are modelled using Gaussian processes. The model presented here is an extension of the model shown in~\cite{Strähnz_arena_24}. The main shower parameters used in this model are the normalisation of the \ldf ($f_0$), the geometric distance to the shower maximum ($\dmax$), the arrival direction ($\theta, \varphi$), and the position in shower plane ($x,y$). These are used with two parametric models to first find the energy fluence, the amount of energy deposited in the ground, in every antenna using the \ldf from~\cite{schlueter_2023_signal_model} which then normalises the spectral shape found with the parametrisation from~\cite{martinelli2023parameterizationofthefrequencyspectrum}. To account for deviations from those parameterisations, two possible sources must be taken into account. First, the scatter of the parametrisations is taken into account individually by adding deviation parameters with a prior distribution according to the spread. Second, the deviation between simulations and measurements is encoded with a Gaussian process, which is added at the electric field level. A non-parametric narrowband \rfi contribution is added as well. Lastly, the instrument response, including the complex antenna gain pattern, is applied to the electric field to find the expected voltage traces for a given set of parameters. These are compared to the measured traces using a Gaussian likelihood.
    
    \subsection{Lateral Distribution of Geomagnetic Emission}
    To model the lateral distribution of the geomagnetic emission, the \ldf from~\cite{schlueter_2023_signal_model} is used. It has the form of the sum of a Gaussian and a Sigmoid (hence the name ``GS'')
    \begin{equation}
        f_\text{GS} = f_0 \left[\exp\left(-\left(\frac{r-r_0^\text{fit}}{\sigma}\right)^{p(r)}\right) + \frac{a_\text{rel}}{1+\exp(s [r/r_0^\text{fit} - r_{02}])} \right],
    \end{equation}
    whereby $r_0, \sigma$ and $p$ describe the Gaussian and $a_\text{rel}, s$ and $r_{02}$ describe the sigmoid. The overall normalisation of the function is set by parameter $f_0$, $r$ is the distance from the shower axis. The shape parameters can be described in terms of the geometric distance to the shower maximum ($\dmax$) as described in Appendix C of~\cite{schlueter_2023_signal_model}. From there it is also evident, that there are large shower to shower fluctuations between these parameters, which must be included as prior information. This is encoded in an extra deviation parameter each, which are added to the shape parameters of $f_\text{GS}$. The prior distribution for these is set as a zero-centred normal distribution with a width corresponding to the spread in the parameterisation.
    
    \subsection{Early-Late Effect and Charge Excess Emission}
    The lateral distribution of the geomagnetic emission can only be described by a rotationally symmetric function if the so-called ``geometric early-late effect'' is corrected. This effect arises because the distance between the emission region (around shower maximum) and the observer is different, depending on whether the observer is placed towards the arrival direction or away from it. The correction factor $c_\text{el}$ depends on the height of the observer above the shower plane ($z_i$) and $\dmax$ and must be applied to the axis distance ($r$) and resulting fluence ($\fluence$) as
    \begin{equation}
        c_\text{el} = 1 + \frac{z_i}{\dmax}, \ \ \ \fluence_\text{symm} = c_\text{el}^2\fluence,\ \ \ r_\text{symm} = \frac{r}{c_\text{el}}.
    \end{equation}
    The charge excess emission can be described by the charge excess fraction $a_\text{ce} = \sin^2(\alpha)\fluence_\text{CE}/\fluence_\text{geo}$, where $\alpha$ is the geomagnetic angle. This can be parametrised as a function of $\dmax$ and the density at shower maximum ($\rho_\text{max}$)~\cite{schlueter_2023_signal_model}. As before, a deviation parameter is added to inform the inference about the scatter in the parametrisation. The resulting fluences at each antenna are thus modeled as:
    \begin{align}
        \fluence_\text{geo} &= \cel^{-2} \fGS(r \cel^{-1}|f_0, \dmax) \label{fgeo}\\
        \fluence_\text{CE}  &= \fluence_\text{geo} \sqrt{a_\text{ce}} \sin^{-2}\alpha\label{fce}
    \end{align}
    
    \subsection{Electric Field, Fluctuations and Likelihood}
    The model for the electric field, as well as for the fluctuations, \rfi, instrument response and likelihood is the same as in~\cite{Strähnz_arena_24}, with the only exception that the amplitude of the electric field is now derived from the fluence as calculated in equations \ref{fgeo} and \ref{fce}. The model for the absolute spectrum of the electric field is taken from~\cite{martinelli2023parameterizationofthefrequencyspectrum}. The phase spectrum is assumed to be linear with a slope determined by the peak time of the trace. 
    Fluctuations and narrowband \rfi are modelled by Gaussian processes. The likelihood is assumed to be a normal distribution with a width equal to the RMS of the data.
    The normalisation of the absolute spectrum of the electric field can be calculated from the energy fluence, which is just the time integral of the Poynting vector.
    For transverse waves, this integral simplifies to $\fluence = \epsilon_0c\int|\vec{E}|^2\text{d}t$.
    Using the Plancherel theorem and knowing that the detector will be band limited between $f_\text{low}$ and $f_\text{up}$, this is equivalent to
    \begin{equation}
        \fluence = \epsilon_0c\int_{-\infty}^{\infty}|\vec{E}|^2\text{d}t = \epsilon_0c\int_{-\infty}^\infty |\vec{\tilde{E}}|^2\text{d}f = \epsilon_0c\int_{f_\text{low}}^{f_\text{up}}|\vec{\tilde{E}}|^2\text{d}f.
    \end{equation}
    If the absolute spectrum of the electric field ($\tilde{E}$) is modelled by an equation with an amplitude parameter ($A$) that follows $\tilde{E}(A, \vec{\xi}) = A \tilde{E}(1, \vec{\xi})$ (with $\vec{\xi}$ representing all other parameters) the spectrum can be normalised to the fluence with:
    \begin{equation}
        \tilde{E}(\fluence,\vec{\xi}) = \sqrt{\frac{\fluence}{\epsilon_0c\int_{f_\text{low}}^{f_\text{up}}|\tilde{E}(1, \vec{\xi})|^2 \text{d}f}} \tilde{E}(1,\vec{\xi})
    \end{equation}
    The integral is evaluated numerically at the point defined by the inference parameters.
    
\section{Performance on Simulations}
    \begin{figure}
        \centering
        \begin{subfigure}[t]{0.49\textwidth}
            \includegraphics[width=\linewidth]{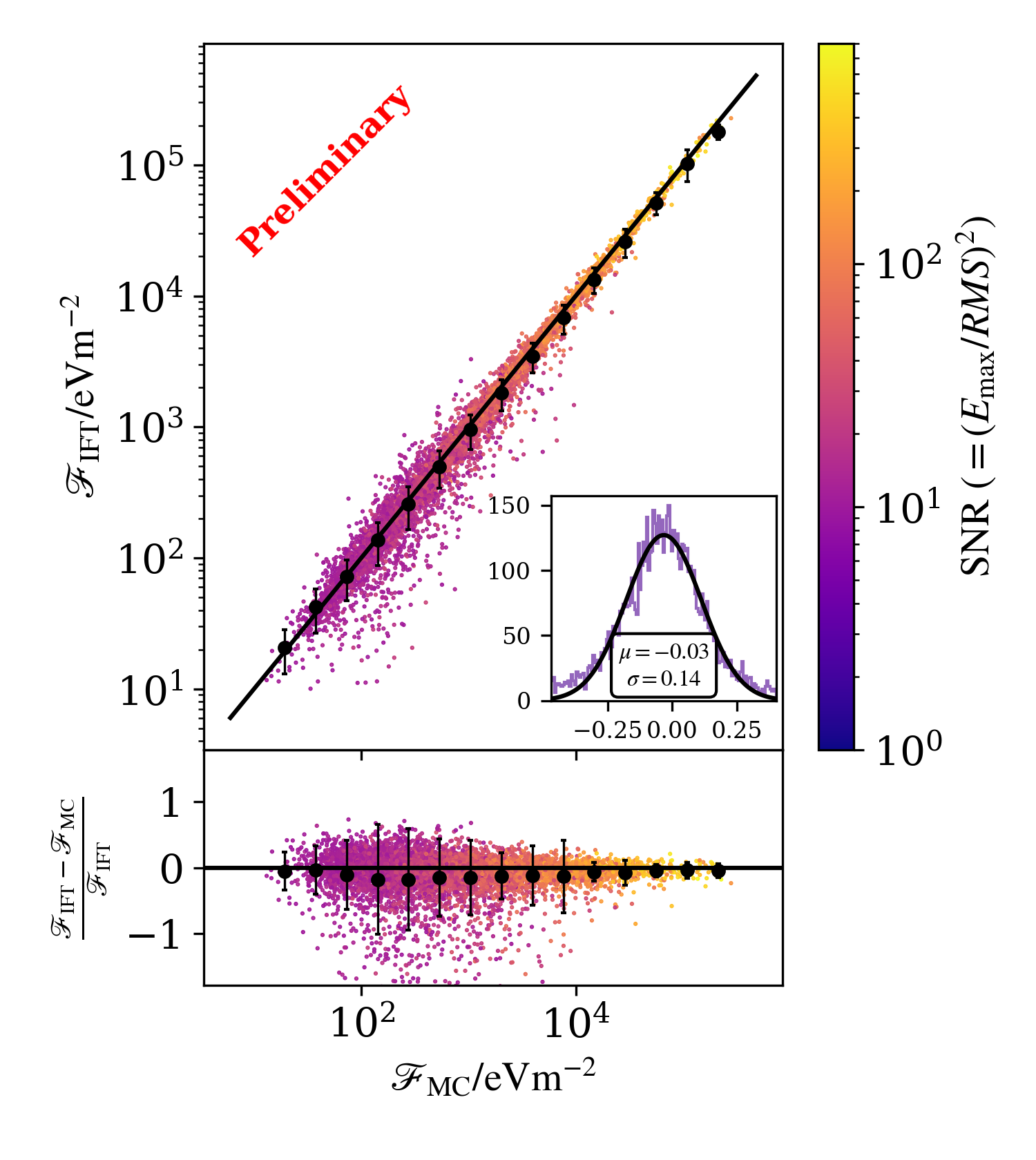}
        \end{subfigure}
        \begin{subfigure}[t]{0.49\textwidth}
            \includegraphics[width=1\linewidth]{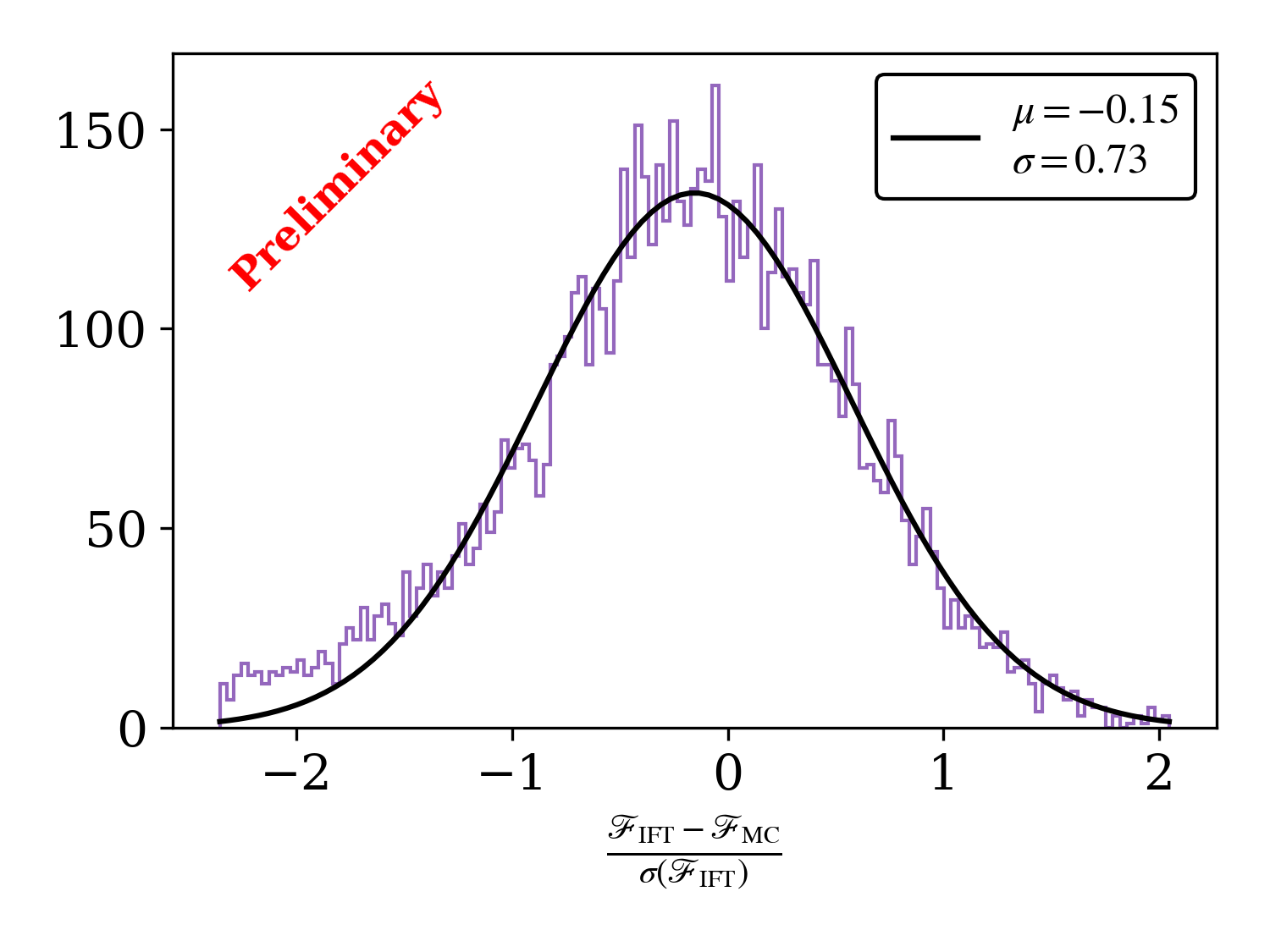}
        \end{subfigure}
        \caption{Comparison of the energy fluence. \emph{Left}: Reconstructed vs.\ simulated energy fluence and their relative residual. The inset shows the distribution of that residual and a Gaussian fit. \emph{Right:} The corresponding pull distribution with a Gaussian fit.}
        \label{fig:fluence}
    \end{figure}
    
    To test this model it has been applied to 2859 CoREAS simulations for inclined proton, helium, nitrogen and iron showers with zenith angles ranging from 65\degree to 85\degree and primary energies from $10^{18.5}$\,eV to $10^{20}$\,eV. They have been forward folded with the detector response of the newly deployed radio detector of the Pierre Auger observatory~\cite{giaccari_2022_antennapattern}. Measured noise has been added at the ADC-count level. They were then reconstructed using the MGVI algorithm with the method described above. Using the extended \geoVI algorithm was tested on a smaller subset, but did not provide sufficiently improved results considering its much longer run time. During reconstruction, only stations with \ansnr larger then 10 where considered. This is necessary, since the integer counts of the (simulated) ADC introduce a discretisation in the (mock-)data. This cannot be correctly captured by a simple Gaussian likelihood, as is currently used in this reconstruction. To select only successful reconstruction, a cut to the reduced $\chi^2 (< 1.05)$ was applied.

    Figure~\ref{fig:fluence} shows a comparison of the reconstructed to the simulated energy fluence, which can be used as a proxy to judge the reconstruction of the electric field. It also shows the pull distribution of the fluence. There is very good agreement between the reconstructed and the simulated fluence. This is to be expected, as the electric field model from~\cite{Strähnz_arena_24} was used, which already showed good agreement for the energy fluence. There is also a small bias towards underestimation. The pull distribution has a standard deviation of 0.73 ($<1$) which indicates an overestimation of the uncertainty.

    \begin{figure}
        \centering
        \begin{subfigure}[t]{0.49\textwidth}
            \includegraphics[width=\linewidth]{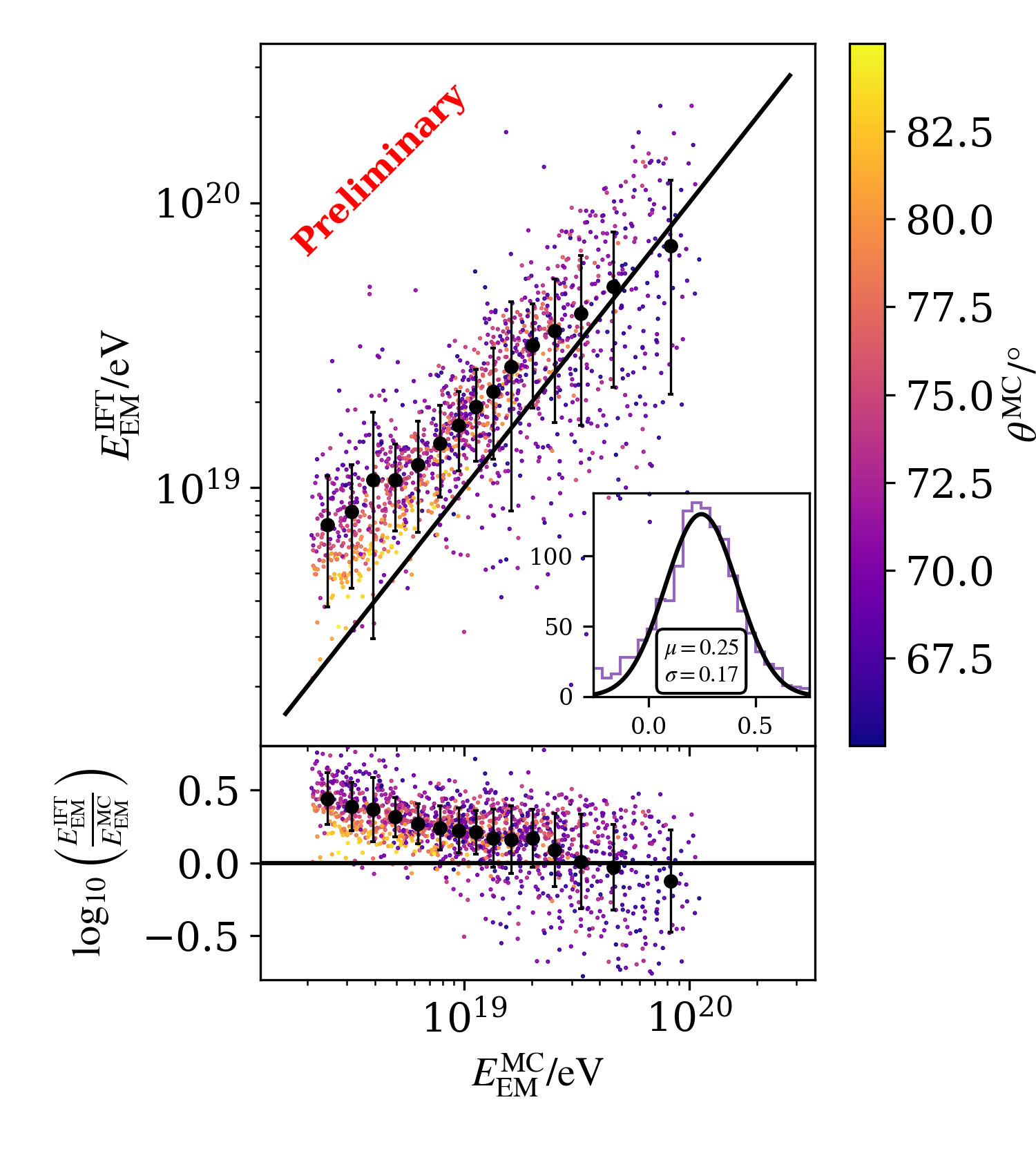}
        \end{subfigure}
        \begin{subfigure}[t]{0.49\textwidth}
            \includegraphics[width=\linewidth]{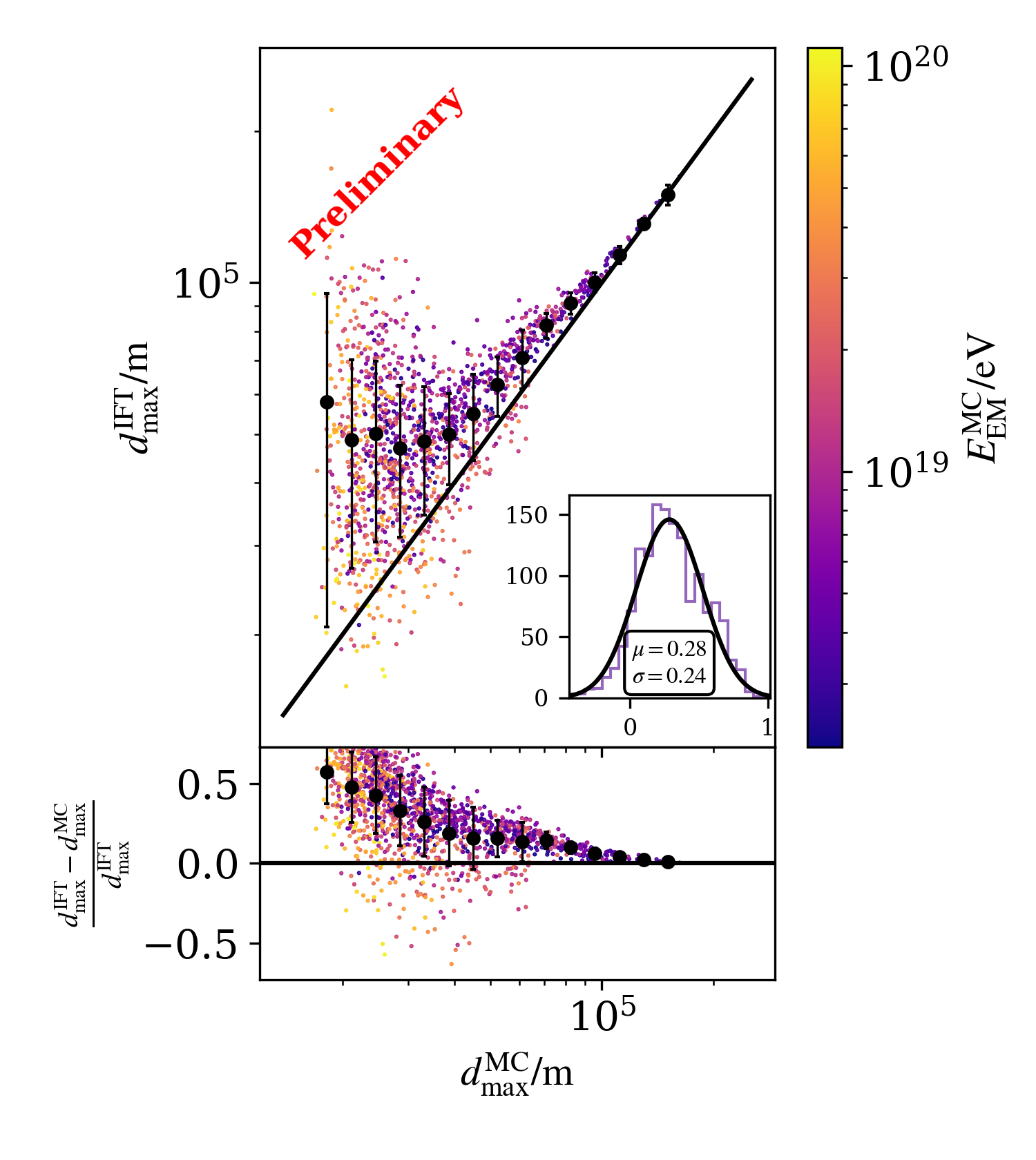}
        \end{subfigure}
        \begin{subfigure}[t]{0.49\textwidth}
            \includegraphics[width=1\linewidth]{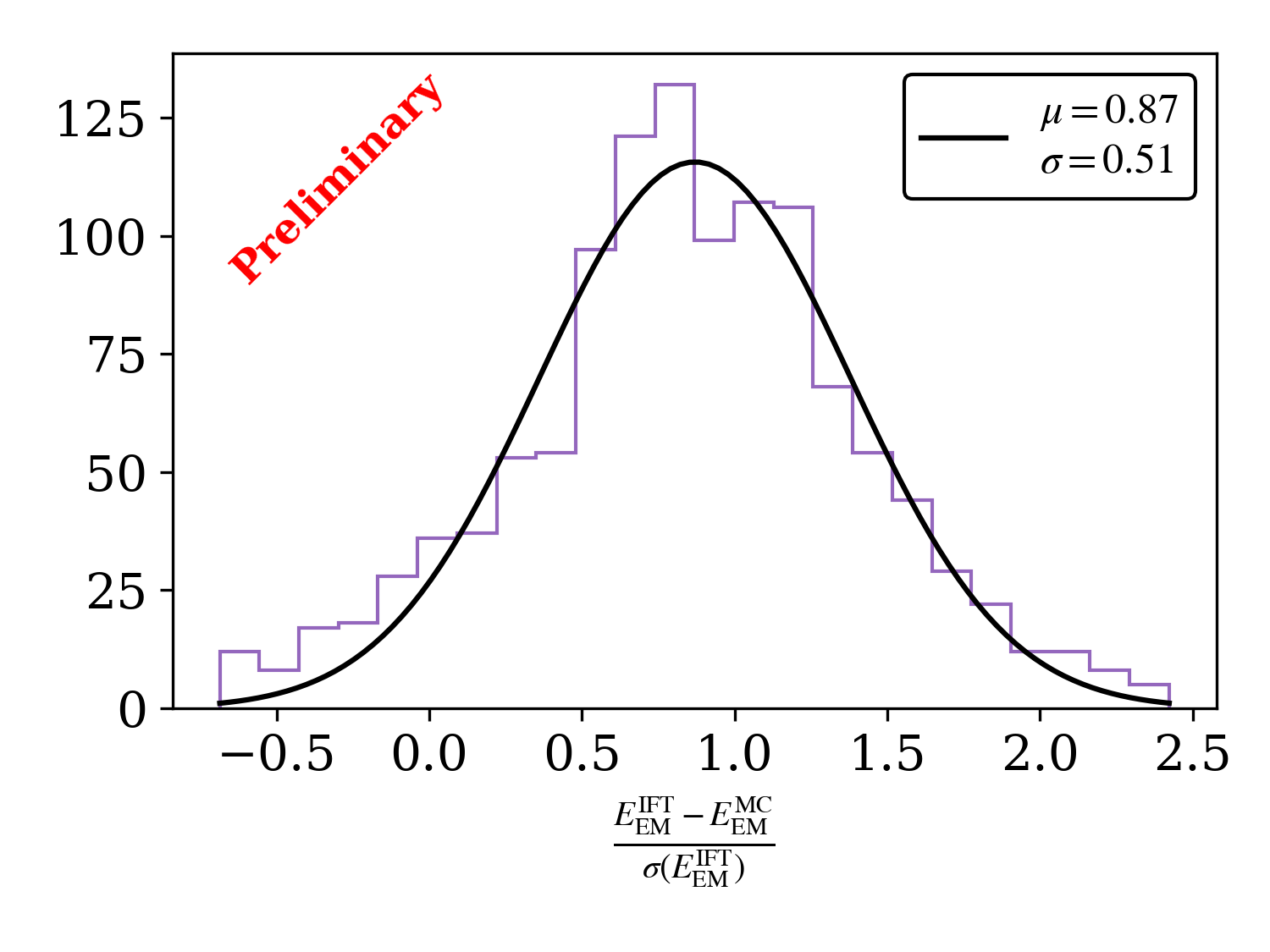}
        \end{subfigure}
        \begin{subfigure}[t]{0.49\textwidth}
            \includegraphics[width=1\linewidth]{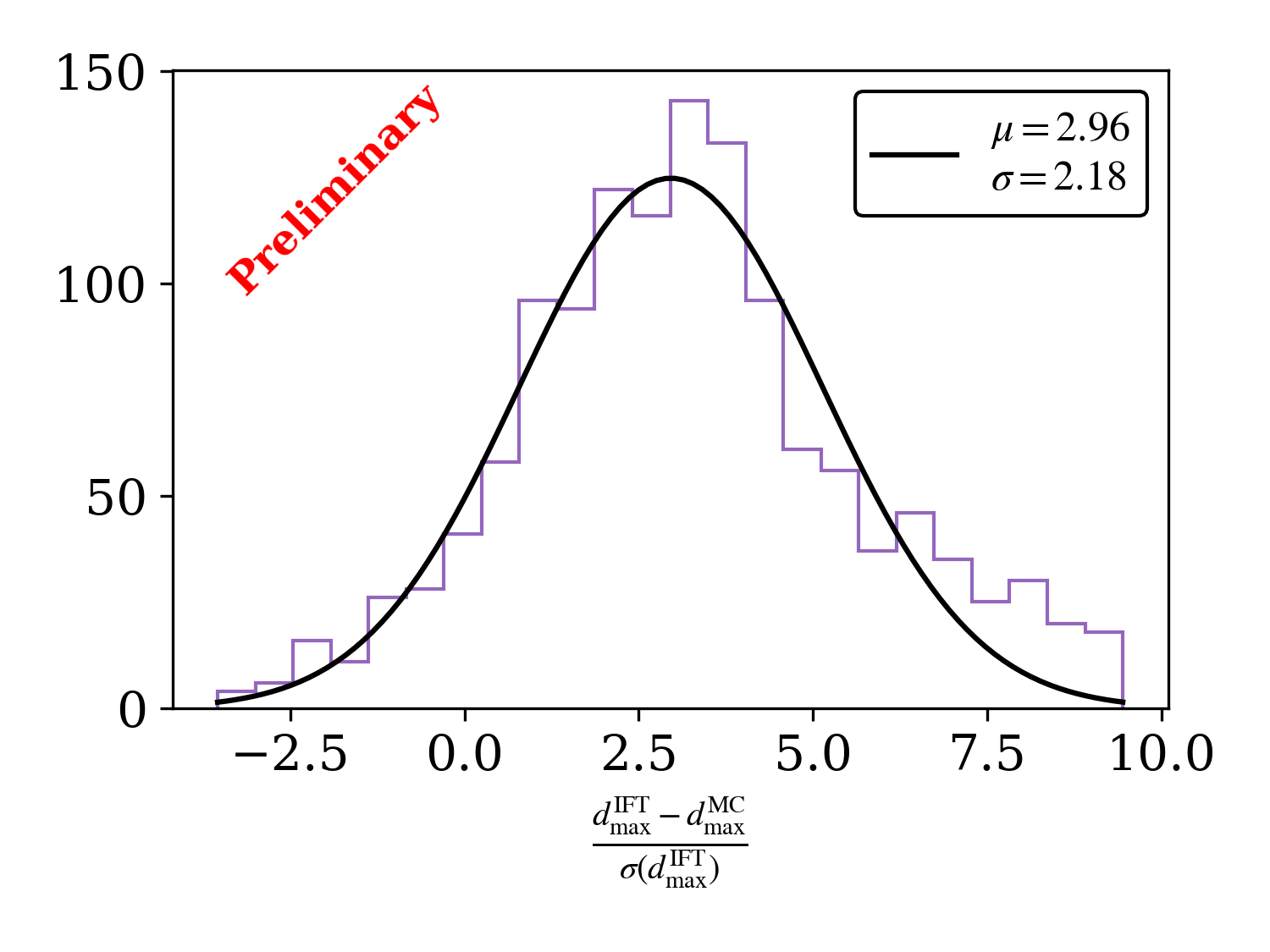}
        \end{subfigure}
        \caption{Comparison of electromagnetic energy and distance to shower maximum between reconstruction and simulation. \emph{Top left}: Reconstructed vs simulated electromagnetic energy and the logarithm of their ratio. The inset shows the distribution of that logarithm with a Gaussian fit. The zenith angle is given as colour. \emph{Bottom left:} The corresponding pull distribution with a Gaussian fit. \emph{Top right:} Reconstructed vs simulated distance to shower maximum and their relative residual. The inset shows the distribution of that residual with a Gaussian fit. The \EM energy is given as colou. \emph{Bottom right:} The corresponding pull distribution with a Gaussian fit.}
        \label{fig:comparison_E_dmax}
    \end{figure}
    \begin{figure}
        \centering
        \begin{subfigure}[t]{0.45\textwidth}
            \includegraphics[width=\linewidth]{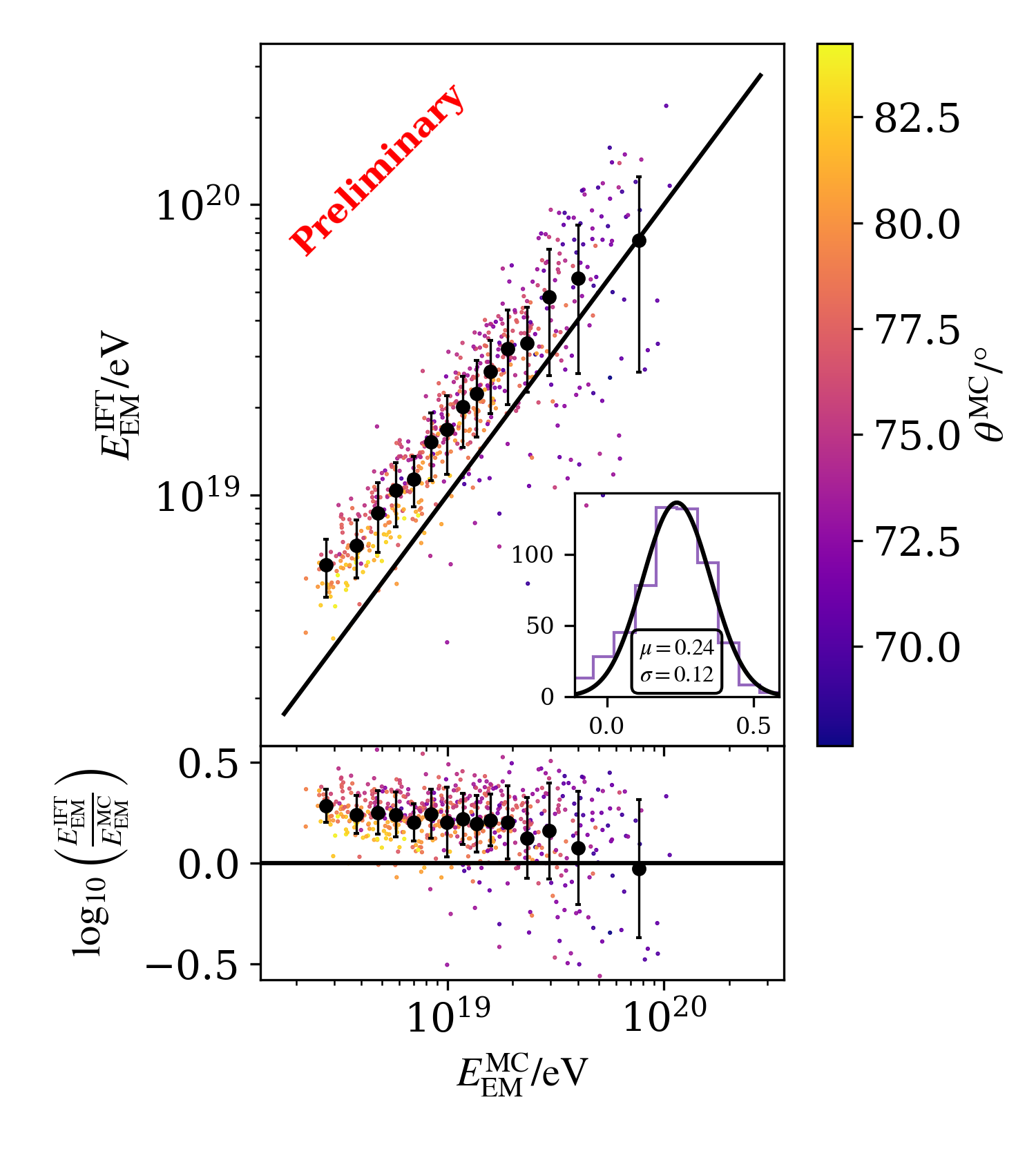}
            \caption{Electromagnetic energy}
        \end{subfigure}
        \begin{subfigure}[t]{0.45\textwidth}
            \includegraphics[width=\linewidth]{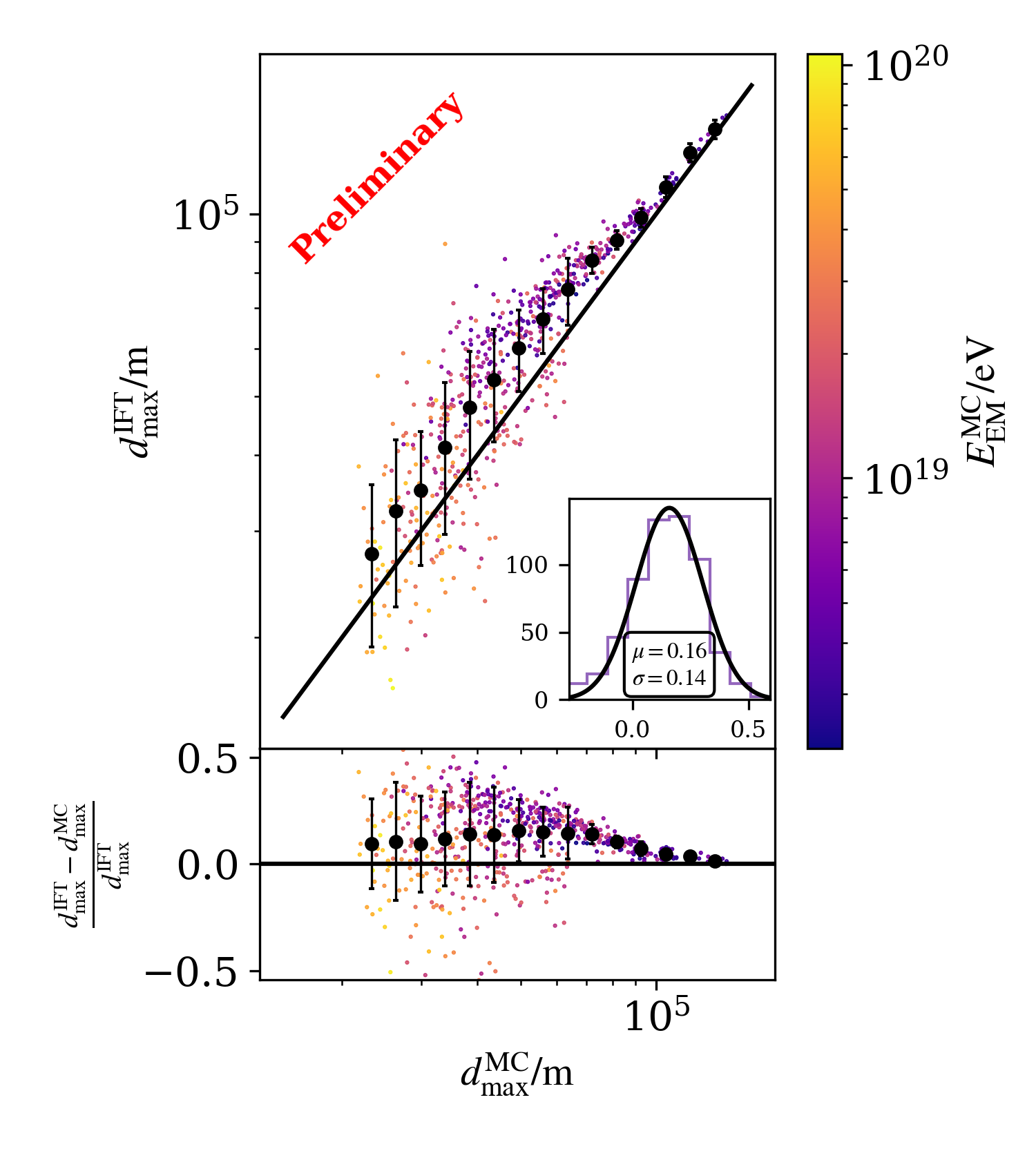}
            \caption{Distance to shower maximum}
        \end{subfigure}
        \begin{subfigure}[t]{0.45\textwidth}
            \includegraphics[width=1\linewidth]{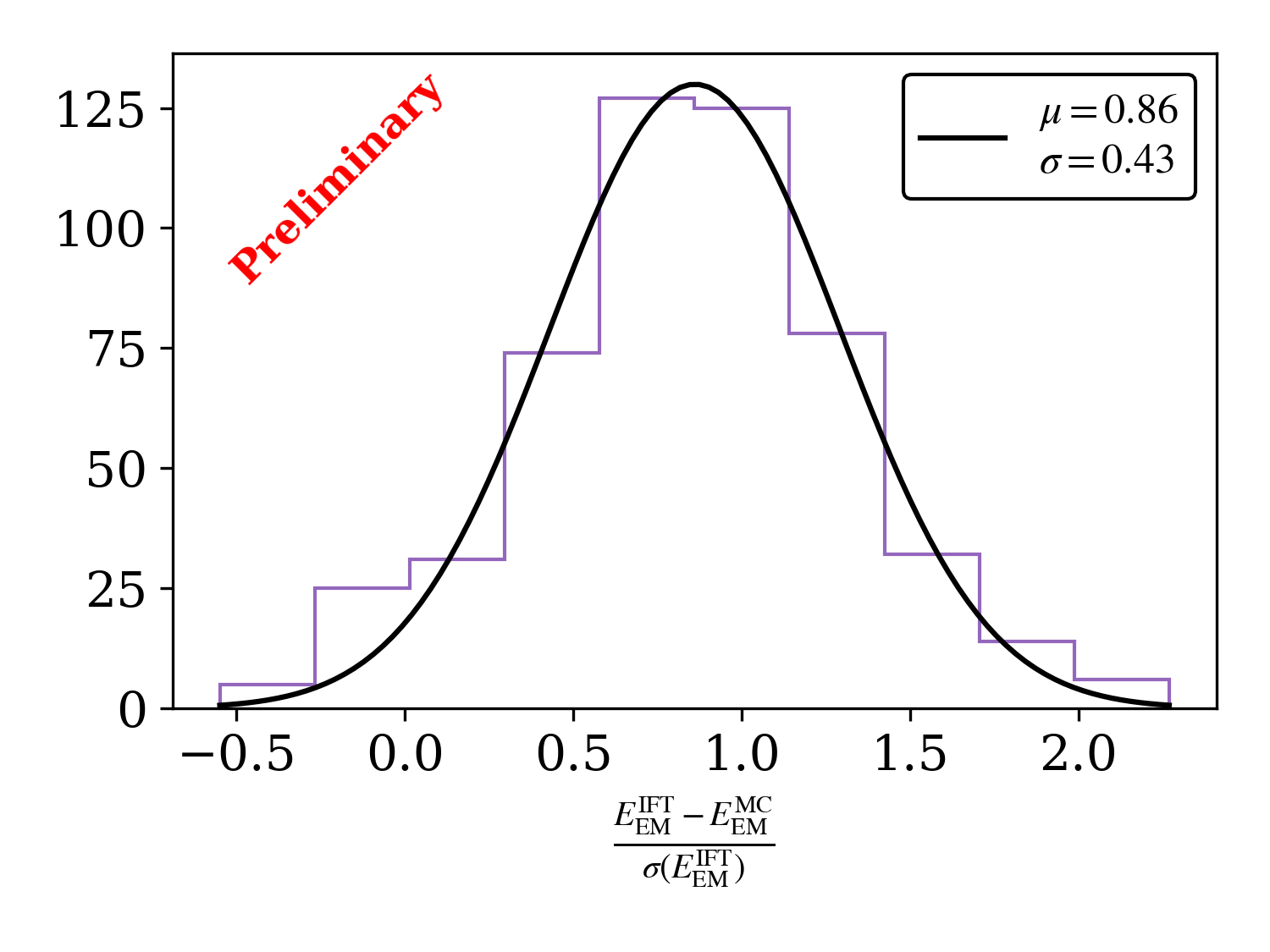}
            \caption{Pull distribution of electromagnetic energy}
        \end{subfigure}
        \begin{subfigure}[t]{0.45\textwidth}
            \includegraphics[width=1\linewidth]{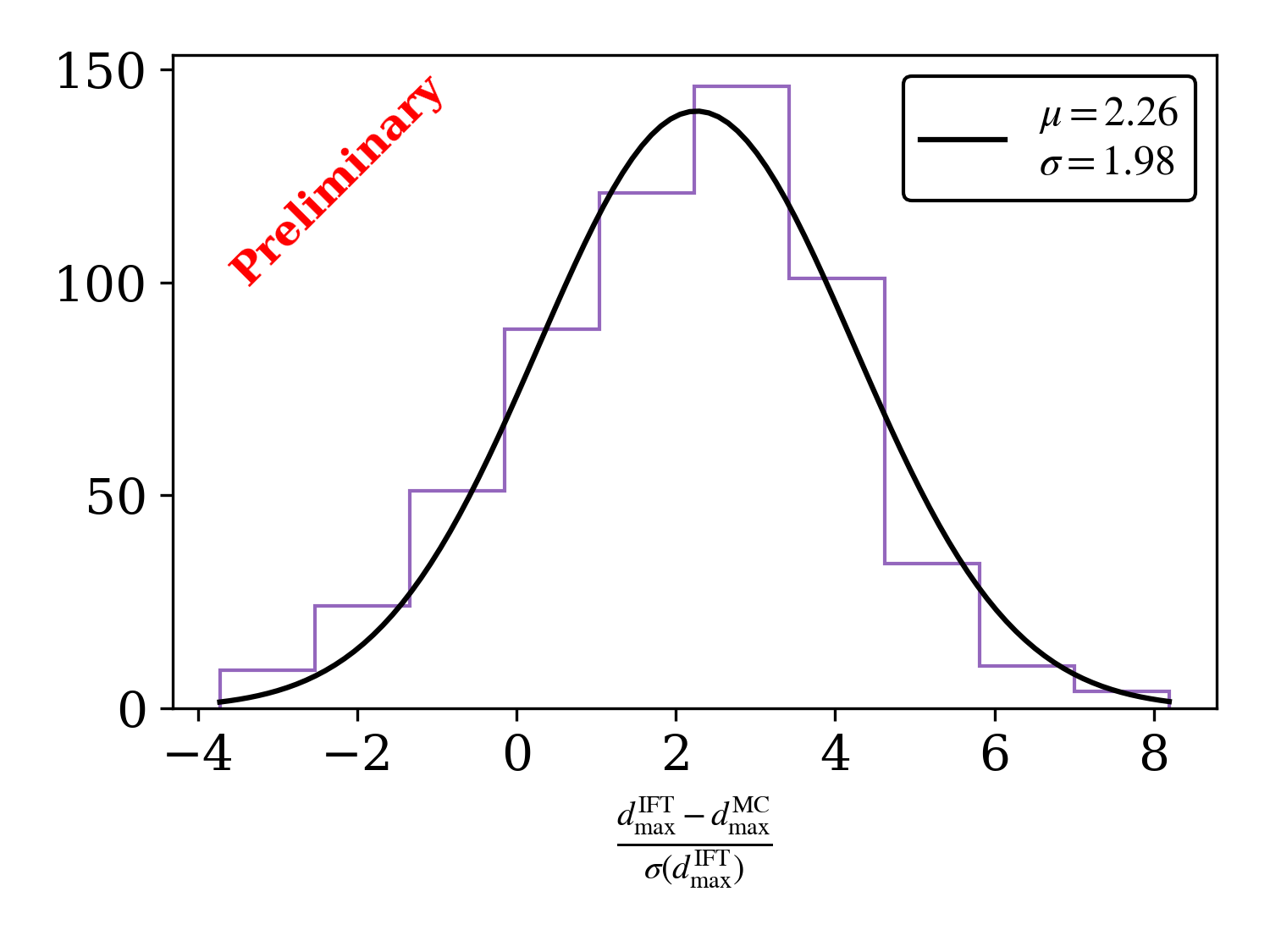}
            \caption{Pull distribution of distance to shower maximum}
        \end{subfigure}
        \caption{Comparison of electromagnetic energy and distance to shower maximum between reconstruction and simulation for events with 5 of more stations. See also~\ref{fig:comparison_E_dmax}}
        \label{fig:comparison_5stations}
    \end{figure}

    Figures~\ref{fig:comparison_E_dmax} and~\ref{fig:comparison_5stations} show comparisons and pull distributions for the \EM energy and the distance to shower maximum for all events and those with at least 5 stations respectively. The agreement for both is noticably worse than for the energy fluence. For all reconstructed events, the \EM energy is biased to overestimation at low energies and has a wide spread. The distance to shower maximum is always overestimated and hardly constrained for deep showers. When considering only showers with 5 or more stations with signal, there is still a bias towards overestimation of the \EM energy, but it is more constant over the entire range, only changing at the highest energies. The spread in $\dmax$ for deep showers is massively reduced. The pull distributions show that the uncertainty of the electromagnetic energy is severely overestimated, while the uncertainty on the distance to the shower maximum is underestimated.

\section{Summary and Discussion}
    We have detailed above a forward model for the reconstruction of inclined air showers using \ift. The model is based on a previously described model for the electric fields of air shower radio emission and adds knowledge about the lateral distribution of signals, aiming to reconstruct the \EM energy and the distance to the shower maximum. The reconstruction of the electric field is solid and comparable to previous work. The reconstruction of the two shower observables still leaves room for improvement.

    The bias in the reconstruction of the energy is to be expected, since the model used to describe the lateral distribution of air shower radio emission was developed for a reconstruction method with energy fluences calculated using the ``noise subtraction'' method~\cite{schlueter_2023_signal_model}. In this method, the fluence is calculated by integrating the electric field in a signal window and then subtracting the integral over a noise window. As in this work, there is no noise in the reconstructed electric field, the fluence is calculated simply as the integral over all time (which corresponds to the integral over all frequencies). This change in definition means that the fluences at every antenna are reported larger than in the dataset the \ldf model was parametrised from. Since the energy is claculated from the integral over the \ldf, higher fluences correspond to higher energies. This also means, that this method is most likely not generally overestimating energies but rather underestimating very large energies. Due to this work-in-progress state, we regard the current results as preliminary. The bad constraining power of $\dmax$ is also expected, as it is only constrained very indirectly through the parametrisations of the \ldf and the slope of the absolute spectrum of the electric field. This will likely improve with the addition of timing information.

\section{Outlook}
    The model shown here is the first attempt at a holistic reconstruction of the radio signals of inclined \eass. Despite the less than optimal performance of the reconstruction of shower observables, it lays the groundwork for a new reconstruction method that promises not only more accurate measurements of the properties of air showers but also better estimates on their uncertainties. In a next step, the biases in the reconstruction, whose causes have been discussed above, must be fully understood and addressed. To complete this holistic model, information about the signal timing must be included, which has turned out to be less trivial than expected, although some promising models are now under development.

\bibliographystyle{JHEP}
\footnotesize
\bibliography{bibliography}

\footnotesize
\section*{Acknowledgements}
Simon Strähnz and Karen Terveer acknowledge funding through the German Federal Ministry of Education and Research for the project ErUM-IFT: Informationsfeldtheorie für Experimente an Großforschungsanlagen (Förderkennzeichen: 05D23EO1).

\end{document}